\documentstyle[12pt,aasms4]{article}

\begin{document}

\title{Integrating the BeppoSAX Gamma-Ray Burst Monitor into the 3rd Interplanetary Network}

\author{K. Hurley}
\affil{University of California, Berkeley, Space Sciences Laboratory,
Berkeley, CA 94720-7450}
\authoremail{khurley@sunspot.ssl.berkeley.edu}

\author{M. Feroci, M.-N. Cinti, E. Costa, B. Preger}
\affil{Istituto di Astrofisica Spaziale - C.N.R., Rome, I-00133 Italy}

\author{F. Frontera, D. Dal Fiume, M. Orlandini, L. Amati}
\affil{ITESRE, Bologna, Italy}

\author{L. Nicastro}
\affil{IFCAI, Palermo, Italy}

\author{J. Heise, J.J.M. in 't Zand}
\affil{Space Research Organization Netherlands, Utrecht, Netherlands}

\author{T. Cline}
\affil{Goddard Space Flight Center, Code 661, Greenbelt, MD, 20771}

\begin{abstract}

\end{abstract}

We have added the \it BeppoSAX \rm Gamma-Ray Burst Monitor to the 3rd
Interplanetary Network of burst detectors.  We analyze 16 bursts whose
positions are known to good accuracy from measurements at other wavelengths.
We show that there is excellent agreement between the \it Ulysses/BeppoSAX \rm
triangulation annuli and the known positions of these events, and that 
these annuli can in many cases provide useful constraints on the positions of
bursts detected by the BeppoSAX Wide Field Camera and Narrow Field Instruments.


\section{Introduction}

It is now well known that the breakthrough in our understanding of cosmic gamma-ray
bursts (GRBs) has come about through the multiwavelength identification of fading
counterparts, and that this revolution was initiated by \it BeppoSAX \rm observations
and precise localizations of X-ray counterparts (e.g., Costa et al. 1997)
with the Wide Field Camera 
(WFC).  Less well known, however, is the fact that \it BeppoSAX \rm
has other ways of localizing bursts, in particular with the Gamma-Ray Burst Monitor
(GRBM).  Here we describe the results of integrating \it BeppoSAX \rm into the 3rd
Interplanetary Network of GRB instruments, and demonstrate that triangulation using
the GRBM and \it Ulysses \rm is capable of producing precise localization annuli whose
accuracy can be verified using the locations of GRB counterparts determined 
from observations at other wavelengths.

\section{Instrumentation}

The \it BeppoSAX \rm GRBM (Feroci et al. 1997; Amati et al. 1997, Frontera et al. 1997)
is the anticoincidence shield for the Phoswich Detection
System.  Briefly, this shield consists of four optically independent CsI(Na) scintillators,
each 1 cm thick, 27.5 cm high, and 41.3 cm wide.  Although the geometric area of
a shield element is 1136 cm$^{2}$, the maximum effective area for a burst with
a typical power law spectrum, arriving at normal incidence, is 
about 420 cm$^{2}$ for units 1 or 3 (the optimum units), when shadowing by
spacecraft structures and the detector housing are taken into account.  The GRBM records data
in either a triggered or a real-time mode, in the energy range $\sim$ 40 - 700 keV.  In the
triggered mode, time resolutions up to 0.48 ms are available; in the real-time mode,
the resolution is 1 s.  For a triggered event, real-time data are also transmitted.  As the spacecraft is in a near-equatorial orbit, it benefits
from a very stable background.  

The \it Ulysses \rm GRB detector (Hurley et al. 1992)  consists of two 3 mm thick
hemispherical CsI(Na) scintillators
with a projected area of about 20 cm$^2$ in any direction.  The detector is mounted
on a magnetometer boom far from the body of the spacecraft, and therefore has a practically
unobstructed view of the full sky.  Because the \it Ulysses \rm mission is in interplanetary
space, the instrument also benefits from an exceptionally stable background.  The
GRB detector operates continuously, and over 97\% of the data are recovered.  The
energy range is $\sim$ 25-150 keV, and like the
GRBM, it takes data in both triggered and real-time modes, with time resolutions as fine
as 8/1024 ms and as coarse as 2 s. Also, like the GRBM, real-time data are transmitted
for triggered events.

\section{Scope of this work}

Because of the very different sensor areas, thicknesses, and energy ranges, it is
not immediately obvious that burst time histories from these two instruments can
be accurately cross-correlated, a necessary prerequisite for precise triangulation.
(The triangulation technique is described in detail in Hurley et al., 1999).
Our goal in this paper is to demonstrate this is indeed the case.  To accomplish
this, we have selected bursts according to the following criteria.

1. The burst must have been observed by both \it Ulysses \rm and the \it BeppoSAX \rm
GRBM, in either triggered or real-time data modes, and

2. The burst must have been independently localized to an accuracy equal to or better
than that achieved by triangulation.  These independent localizations include not only
WFC observations, but also \it BeppoSAX \rm Narrow Field Instrument (NFI) pointings,
optical and radio counterparts, and \it Rossi \rm X-Ray Timing (RXTE) All Sky Monitor
detections.

Many of the bursts which satisfy these criteria have also been observed by other
GRB instruments, notably BATSE and Konus (see table 1).  In some, 
but not all cases, BATSE/\it Ulysses \rm
triangulation will result in somewhat more precise localizations than those presented
here due to BATSE's larger area and better statistics.  We defer these results to another paper.

\section{Observations and Results}

Table 1 lists the dates, times, \it Ulysses \rm and GRBM data modes, and the other experiments
which observed the bursts.  Table 2 gives the right ascension and declination of
the centers of the IPN annuli, their radii, and their 3 $\sigma$ half-widths.  For those bursts
for which an unambiguous counterpart was identified with greater precision than
the annulus width, the last column in the table gives the angular distance $\Delta$ between
the counterpart and the centerline of the annulus , expressed in units of the
number of $\sigma$ of the annulus width.  Figures 1-16 show the localization maps. 

Figure 1 shows the IPN annulus and the BeppoSAX WFC error circle (Heise et al.,
1998) for GRB970111.  (All the WFC error circles in this paper are 99\%
confidence, or 2.58 $\sigma$ regions.)  The NFI error circle indicates the
position of a weak X-ray source.  Since the source lies outside the annulus, this
localization supports the hypothesis that the X-ray source and the burst
are unrelated (Feroci et al. 1998).  Two radio sources were detected in the vicinity
of the WFC error box (Galama et al. 1997), one of which is shown in the figure (the
other lies outside the boundaries).  As neither displayed any fading behavior, they
are not considered to be counterpart candidates.  No fading optical sources were detected
either.

Figure 2 shows the BeppoSAX WFC and NFI error circles for GRB970228 (Heise et al., 1998; Costa et al., 1997).  The IPN annulus was obtained from reprocessed \it Ulysses
\rm data, and differs very slightly from the one in Hurley et al. (1997).  The position of
the fading optical counterpart is also shown (van Paradijs et al. 1997).

Figure 3 shows the IPN annulus for GRB970402.  This burst was quite weak, and was only
detected in the real-time, low resolution data mode of \it Ulysses \rm; although the
GRBM triggered on it, the trigger occurred late in the event and we have used the real-time data. 
A fading X-ray source was found in a BeppoSAX NFI observation (Nicastro et al. 1998a), but
no optical counterpart was ever detected.  \it Ulysses \rm and GRBM time histories of
this event are shown in figure 17.

Figure 4 shows the IPN annulus, and the BeppoSAX WFC and NFI error circles for GRB970508 
(Heise et al., 1998; Piro et al. 1998a).  The position of the optical
counterpart is also indicated (Djorgovski et al. 1997).  This burst too was quite weak,
and although it triggered the BeppoSAX GRBM, it was only observed in the real-time
data of \it Ulysses \rm.  

Figure 5 gives the IPN annulus and the RXTE error box for GRB970815 (Smith et al. 1999).
The RXTE error box is from the All Sky Monitor, and is defined by the response functions 
of two crossed collimators; the box therefore has equal probability per unit area everywhere.
A slightly more precise IPN annulus, derived from BATSE and \it Ulysses \rm, appears
in Smith et al. (1999).

Figure 6 shows the IPN annulus, and the BeppoSAX WFC (Heise et al. 1997) and NFI
(Antonelli et al. 1997) error circles for GRB971214.  The position of the optical 
counterpart (Halpern et al. 1998) is also shown.  IPN and RXTE locations appeared
in Kippen et al. (1997).

Figure 7 gives the IPN annulus and the BeppoSAX WFC (Coletta et al. 1997) error
circle for GRB971227.  The burst was weak and was detected only marginally by \it Ulysses \rm, 
and was detected in the real-time data of the GRBM.  Consequently
the annulus is subject to rather large systematic uncertainties.  A weak X-ray source
detected at the 4 $\sigma$ level in an NFI observation has been proposed as the fading
X-ray counterpart by Piro et al. (1997) and Antonelli et al. (1999).  No radio or optical counterpart was identified.  The WFC error box is large due to poor attitude
reconstruction for this event.

Figure 8 gives the IPN annulus and the BeppoSAX WFC (in 't Zand et al. 1998a) error circle for
GRB980109; no NFI observation was carried out, due to the relatively large uncertainty
in the WFC localization (again due to poor attitude reconstruction).  A possible optical counterpart was 
initially identified, but is no longer considered to be related to the GRB (E. Pian and T. Galama, private
communication).  It lay within the preliminary IPN annulus, but it lies just outside the
final one.

Figure 9 shows the IPN annulus and the BeppoSAX WFC error circle (Celidonio et al. 1998)
for GRB980326.  No NFI observations were carried out, but Groot et al. (1997) identified
an optical transient in the WFC error circle.  A preliminary IPN annulus appeared in
Hurley (1998a).

Figure 10 gives the IPN annulus, and the BeppoSAX WFC (Frontera et al. 1998) and NFI
(in 't Zand et al. 1998b) error circles for GRB980329.   
A radio counterpart was identified by Taylor et al. (1998), and an optical
counterpart was found at the same position by Palazzi et al. (1998).  A preliminary IPN
annulus appeared in Hurley (1998b).

Figure 11 shows the IPN annulus, and the BeppoSAX WFC (Soffita et al. 1998) and 
the two revised NFI
(Piro et al. 1998b) error circles for GRB980425. The position of source 1 is consistent
with that of the unusual supernova 1998bw (Galama et al. 1998a).  However, because the
burst was weak, it was detected only in the \it Ulysses \rm real time data, and the IPN
annulus is wide; it cannot be used to determine which NFI source is associated with
the GRB.   

Figure 12 shows the IPN annulus, the BeppoSAX WFC (Muller et al. 1998) and
NFI (Nicastro et al. 1998b) error circles,
and the position of the optical transient (Hjorth 1998) for GRB980519.

Figure 13 shows the IPN annulus, the RXTE error box (Smith et al. 1999), the
NFI source location (Vreeswijk et al. 1999) and the position of the optical and radio counterparts (Bloom
et al. 1998) for GRB980703.  This annulus is consistent with, but narrower than, the
initial BATSE/Ulysses annulus (Hurley and Kouveliotou 1998). As for GRB970815, 
the RXTE error box is from the All Sky Monitor, and is defined by the response functions 
of two crossed collimators; the box therefore has equal probability per unit area everywhere.

Figure 14 shows the IPN annulus and the RXTE/ASM error box (Smith et al. 1999) for GRB981220.
A radio source was proposed as a possible counterpart (Galama et al. 1998b, Frail \&
Kulkarni 1998), but is no longer thought to be related to the GRB (Frail, Kulkarni \&
Taylor 1999; Hurley \& Feroci 1999).  A preliminary annulus has appeared in
Hurley et al. (1999).

Figure 15 shows the IPN annulus, the BeppoSAX WFC (Feroci et al. 1999) and NFI (Heise
et al. 1999) error circles, and the position of the optical transient
(Akerlof 1999) for GRB990123.  A preliminary IPN annulus was circulated in
Hurley (1999).

Figure 16 shows the IPN annulus, the BeppoSAX WFC (Dadina et al. 1999) and NFI
(Kuulkers et al. 1999) error circles, and the position of the optical transient
(Galama et al. 1999) for GRB990510.

We have selected the light curves of two bursts which were not observed by
BATSE for display in figures 17 and 18.  These show the GRBM and Ulysses real-time and triggered
data.  \section{Discussion}

The \it BeppoSAX \rm GRBM is clearly a sensitive burst detector which makes
an important contribution to the IPN by providing high quality data for events
which are not observed by BATSE.  This is the case for three of the bursts discussed
here.  (This number is fewer than would have been predicted,
based on a probability of 38\% that BATSE
will detect any burst above its threshold - Meegan et al. 1996; however, some
of the events in this paper were in effect selected because of the knowledge of
their detection by BATSE.)  We have demonstrated
that, despite the very different properties of the GRBM and the \it Ulysses \rm
GRB experiment, very accurate triangulations can be done.  In the case of bright
bursts, these annuli can be used to reduce or further constrain the WFC and
NFI error boxes.

\acknowledgments
KH is grateful to JPL for Ulysses support under Contract 958056,
and to the NASA Astrophysics Data Program for supporting the integration
of BeppoSAX into the IPN under NAG5-7766.  BeppoSAX is a program of the
Italian Space Agency, with participation of NIVR, the Dutch Space Agency.

\newpage

\figcaption{IPN annulus and the BeppoSAX WFC and NFI error boxes for GRB970111.
The position of one of the radio sources reported by Galama et al. (1997) is also
indicated. \label{fig1}}

\figcaption{IPN annulus and the BeppoSAX WFC and NFI error boxes for GRB970228.
The position of the optical transient (van Paradijs et al. 1997) is also shown. \label{fig2}
}

\figcaption{IPN annulus and the BeppoSAX NFI source for GRB970402. \label{fig3}
}

\figcaption{IPN annulus, BeppoSAX WFC and NFI error circles, and optical transient
location for GRB970508. \label{fig4}
}

\figcaption{IPN annulus and RXTE error box for GRB970815. \label{fig5}
}

\figcaption{IPN annulus, BeppoSAX WFC and NFI error circles, and optical transient
location for GRB971214. \label{fig6}
}

\figcaption{IPN annulus, and BeppoSAX WFC and NFI error circles for GRB971227. \label{fig7}
}

\figcaption{IPN annulus, BeppoSAX WFC error circle, and possible optical counterpart
of GRB980109. \label{fig8}
}

\figcaption{IPN annulus, BeppoSAX WFC error circle, and optical transient location
for GRB980326. \label{fig9}
}

\figcaption{IPN annulus, BeppoSAX WFC and NFI error circles, and radio counterpart
location for GRB980329.  The optical counterpart location is the same as that
of the radio counterpart. \label{fig10}
}

\figcaption{IPN annulus (upper left and lower right hand corners), and the BeppoSAX
WFC and NFI error circles for GRB980425.  NFI source 1 is associated with SN1998bw.
\label{fig11}
}

\figcaption{IPN annulus, BeppoSAX WFC and NFI error circles, and optical transient position
for GRB980519.
\label{fig12}
}

\figcaption{IPN annulus, RXTE error box, BeppoSAX NFI error circle, and position
of the optical transient and radio counterpart of GRB980703.
\label{fig13}
}

\figcaption{IPN annulus, RXTE error box, and position of radio source for GRB981220.  
\label{fig14}
}

\figcaption{IPN annulus, BeppoSAX WFC and NFI error circles, and optical
transient position for GRB990123.
\label{fig15}
}

\figcaption{IPN annulus, BeppoSAX WFC and NFI error circles, and optical
transient position for GRB990510.
\label{fig16}
}

\figcaption{\it Ulysses \rm (top) and GRBM (bottom) time histories of
GRB970402.  Both instruments observed the burst in real-time data modes.
The \it Ulysses \rm data have been regrouped to 25 s resolution
to improve statistics.  Dashed lines indicate the background levels.
\label{fig17}
}

\figcaption{\it Ulysses \rm (top) and GRBM (bottom) time histories of GRB 981220.
Both instruments observed this event in triggered data modes.  Dashed lines
indicate the background levels.
\label{fig18}
}

\clearpage
\begin{deluxetable}{ccccc}
\tablecaption{\it BeppoSAX GRBM/Ulysses gamma-ray bursts.  }
\tablehead{
\colhead{Date} & \colhead{UT, s}  & \colhead{\it Ulysses \rm data} & \colhead{GRBM data} & \colhead{Other observations \tablenotemark{1}} 
}
\startdata
970111 	&	35040 & T\tablenotemark{2} & T & BATSE \#5773, Konus, DMSP, WFC	\nl
970228	&	10681	& T			   & T & Konus, WFC                    \nl
970402	&     80379 & T & R\tablenotemark{3} & Konus, WFC                    \nl
970508	&     78050 & R			   & T & BATSE \#6225, Konus, WFC       \nl
970815	&	43624	& R			   & T & BATSE \#6335, Konus, RXTE \nl
971214	&	84041 & R			   & T & BATSE \#6533, Konus, NEAR, RXTE, WFC \nl
971227      &     30187	& R			   & R & BATSE \#6546, Konus, WFC       \nl
980109    	& 	04346	& R			   & R & BATSE \#6564, Konus, WFC	 \nl
980326	&	76733 & R			   & T & BATSE \#6660, Konus, WFC       \nl
980329	&	13478 & T			   & T & BATSE \#6665, COMPTEL, Konus, WFC \nl
980425      &	78549 & R			   & T & BATSE \#6707, Konus, WFC       \nl
980519 	&	44412	& R			   & R & BATSE \#6764, Konus, WFC	\nl
980703 	&	15765	& R			   & T & BATSE \#6891, RXTE	\nl
981220 	&	78752	& T			   & T & Konus, RXTE		\nl
990123 	&	35216	& R			   & T & BATSE \#7343, COMPTEL, Konus, WFC \nl
990510 	&	31746	& R			   & R & BATSE \#7560, Konus, WFC		\nl

\enddata
\tablenotetext{1}{Konus: experiment aboard the Wind spacecraft. DMSP: Defense Meteorological Satellite Program.  RXTE: \it Rossi \rm X-Ray Timing Explorer.  NEAR: Near Earth Asteroid Rendezvous mission.}
\tablenotetext{2}{Recorded in triggered mode}
\tablenotetext{3}{Recorded in untriggered (real-time) mode}
\end{deluxetable}

\clearpage
\begin{deluxetable}{cccccc}
\tablecaption{IPN annuli.  }
\tablehead{
\colhead{Date} & \colhead{$\alpha$ \rm (2000)}  & \colhead{$\delta$ \rm (2000)} &
\colhead{Radius} & \colhead{$\delta$ \rm R} & \colhead{$\Delta, \sigma$} \\
\colhead{} & \colhead{(deg.)} & \colhead{(deg.)} & \colhead{(deg.)} & \colhead{(deg.)} 
}
\startdata
970111 	&	177.732 & +33.366 & 49.996 & 0.029	&       \nl
970228	&	165.945 & +36.526 & 83.423 & 0.009  & 2.05  \nl
970402      &     336.646 & -35.576 & 64.726 & 2.694  & 0.006 \nl
970508	&     151.909 & +32.056 & 51.227 & 0.145  & 0.20  \nl
970815	&	159.651 & +20.562 & 68.549 & 0.040  &       \nl
971214	&     171.330 & +11.721 & 53.744 & 0.041  & 0.36  \nl
971227      &     170.783 & +11.569 & 50.859 & 0.119  & 2.5   \nl
980109      &     349.654 & -11.593 & 52.763 & 0.091  &       \nl
970326	&	154.714 & +13.166 & 40.778 & 0.043  & 2.1   \nl
970329	&     154.299 & +13.170 & 49.905 & 0.016  & 0.64  \nl
980425	&     330.566 & -12.778 & 49.670 & 0.428  & 0.30  \nl
980519      &     329.485 & -11.874 & 89.982 & 0.043  & 1.2  \nl
980703      &	331.648 & -9.066  & 33.076 & 0.070  & 1.6   \nl
981220 	&     347.398 & +7.338  & 67.120 & 0.005  &       \nl
990123      &	163.720 & -9.464  & 81.346 & 0.010  & 0.48 \nl
990510      &     144.800 & -7.236  & 78.073 & 0.020  & 1.1  \nl

\enddata

\end{deluxetable}

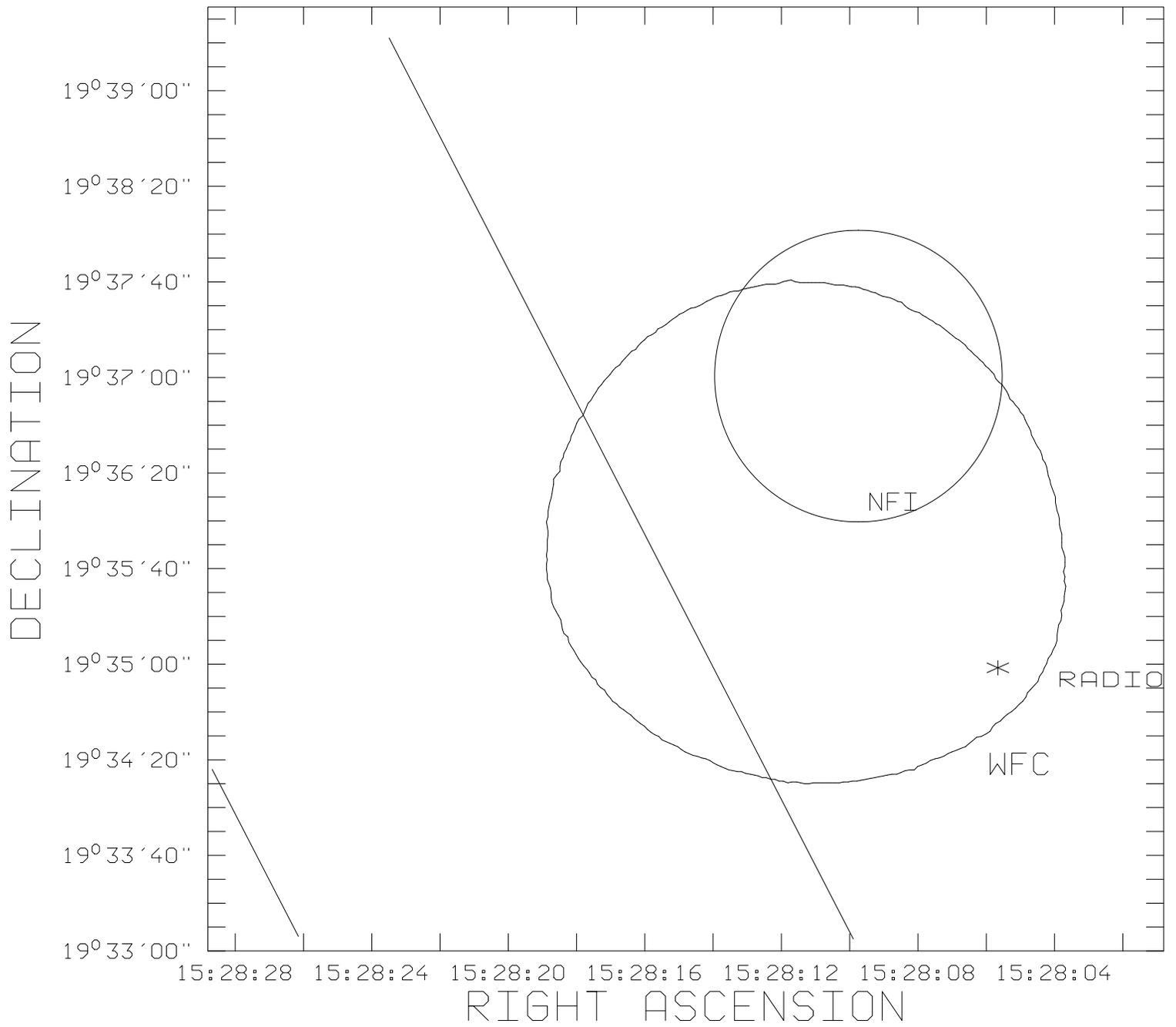
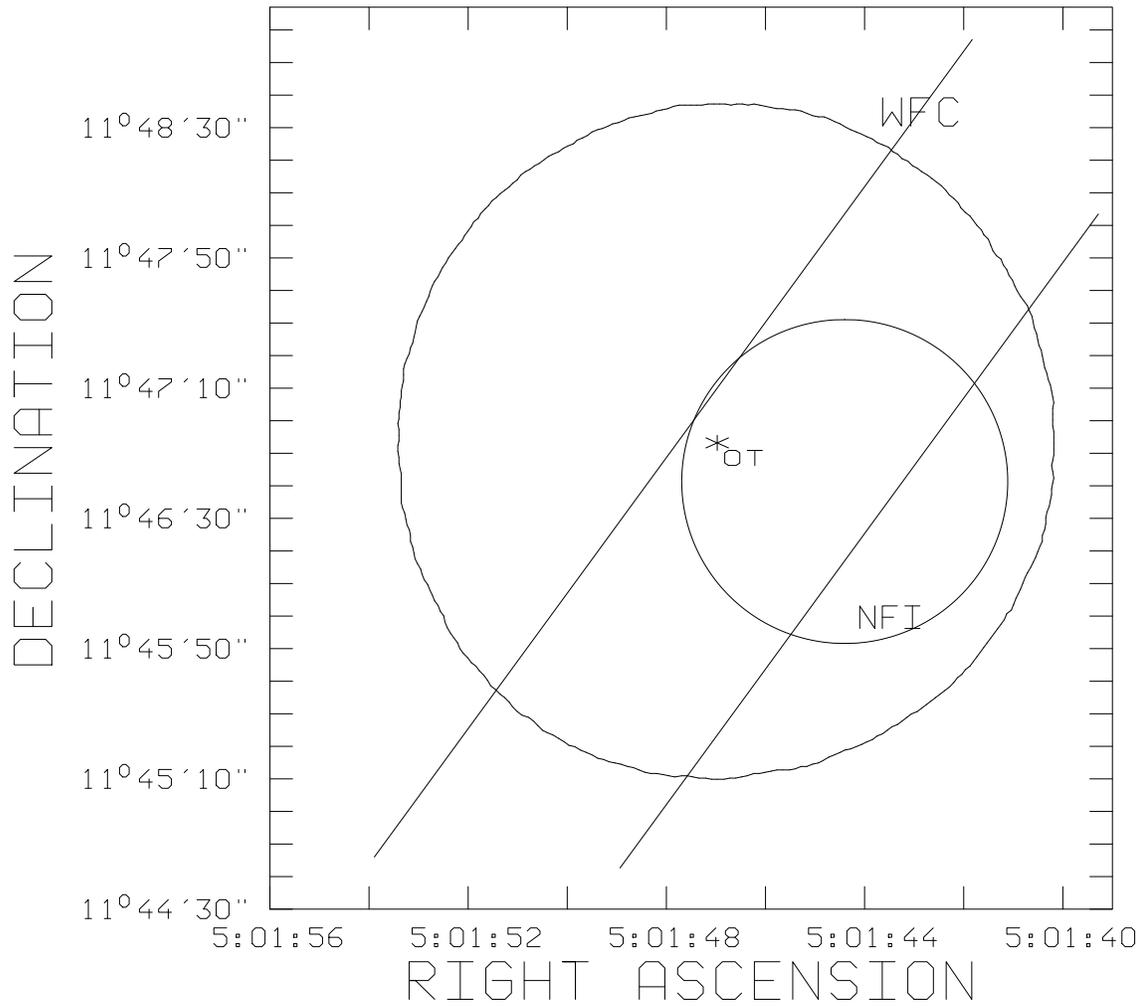
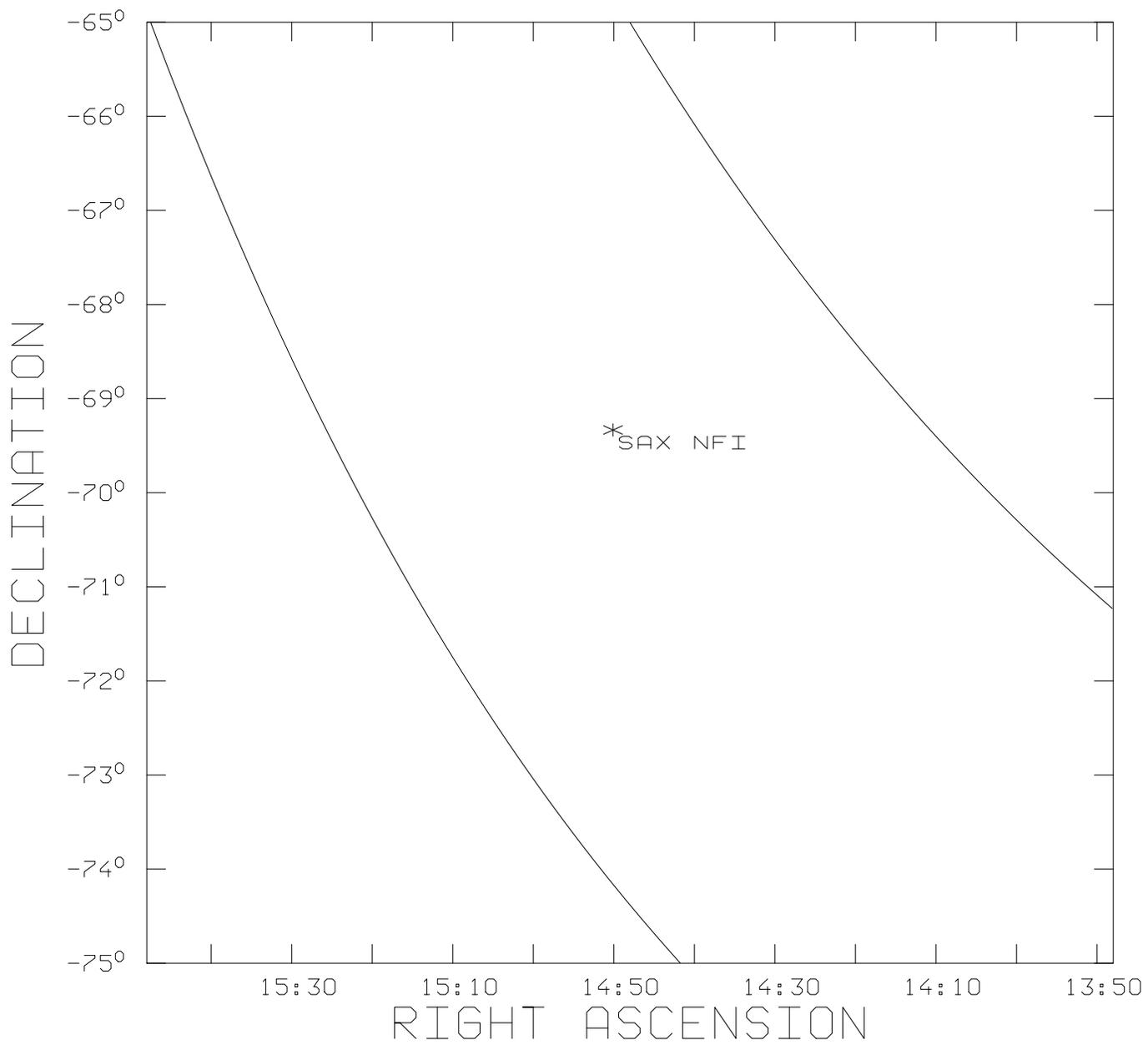
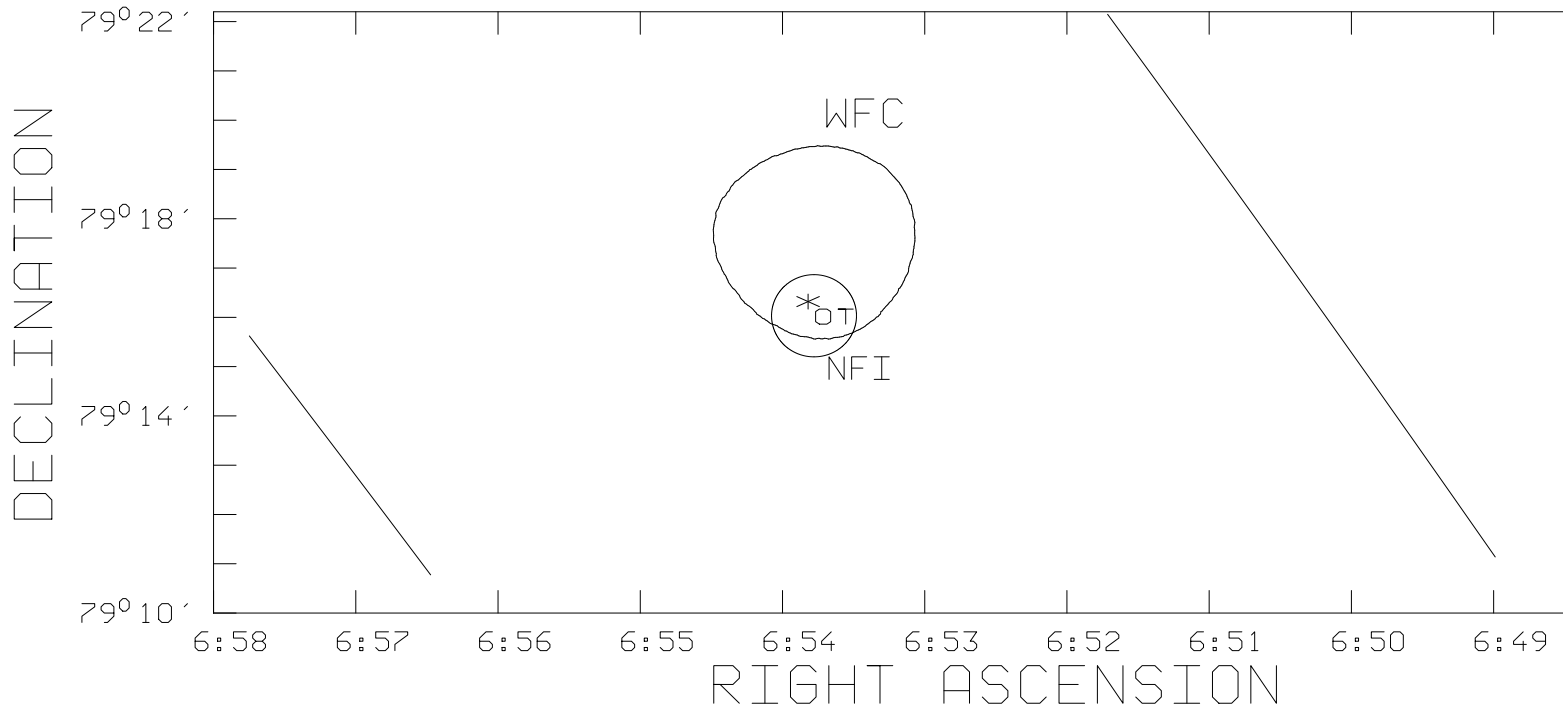
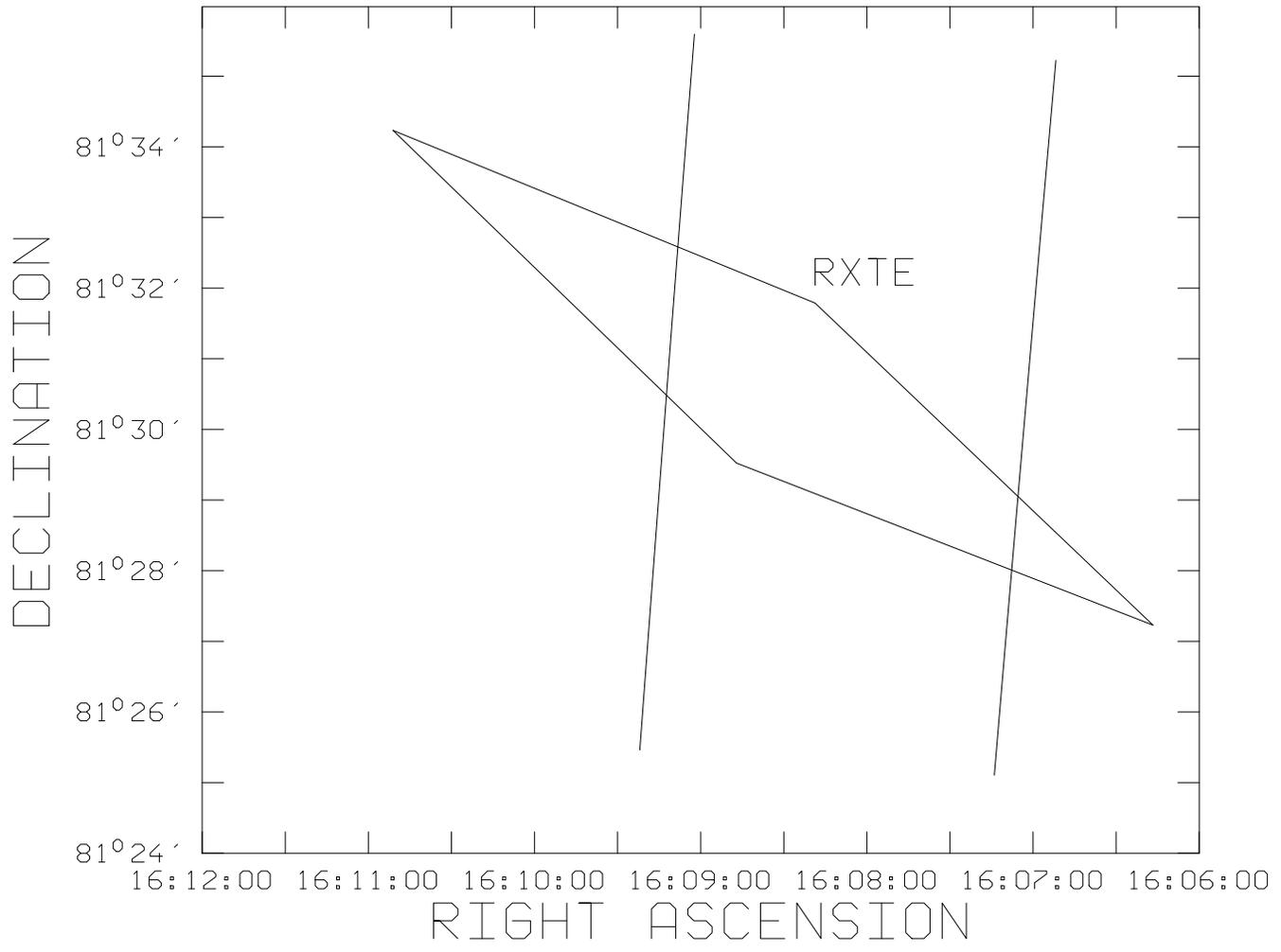
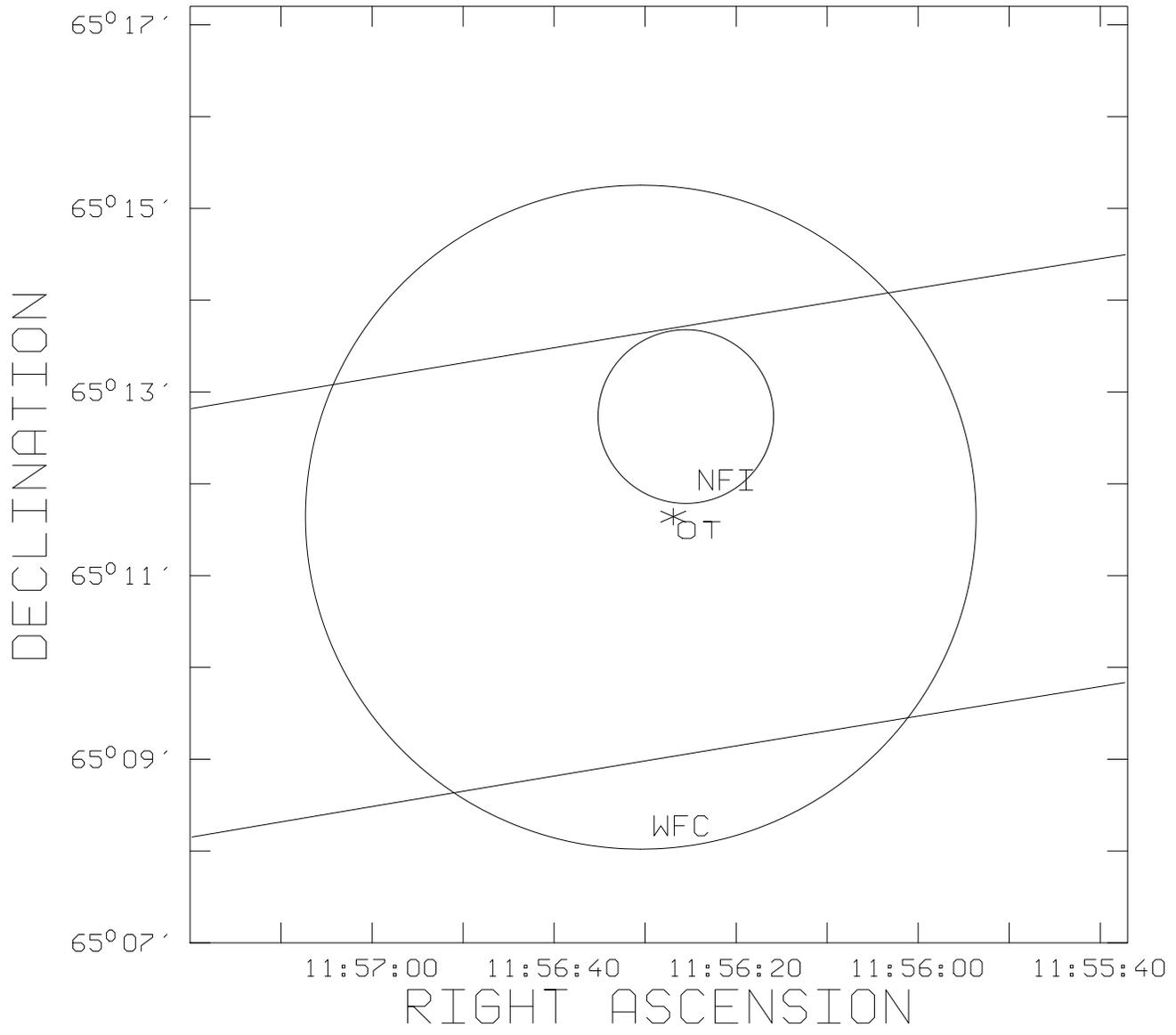
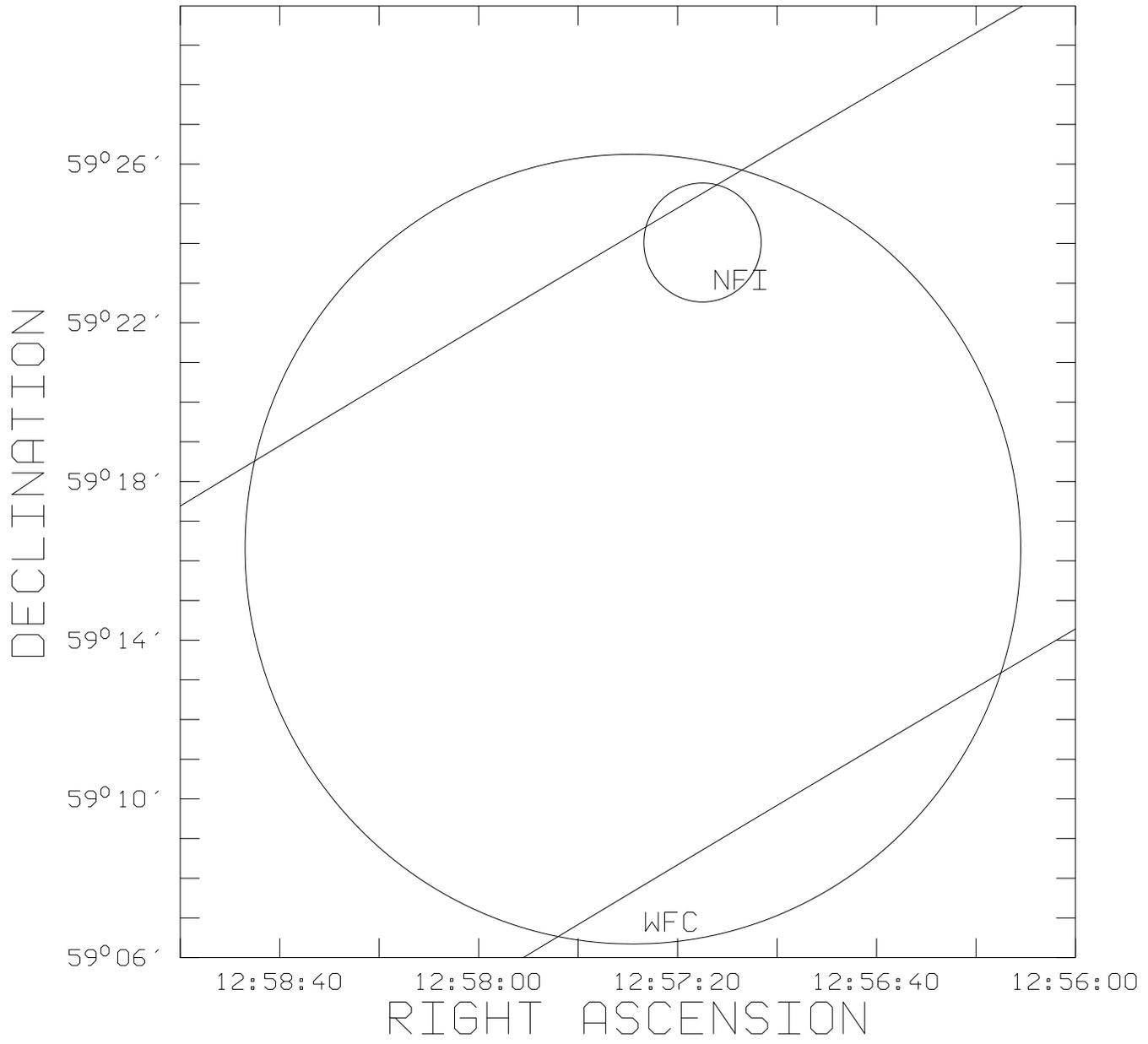
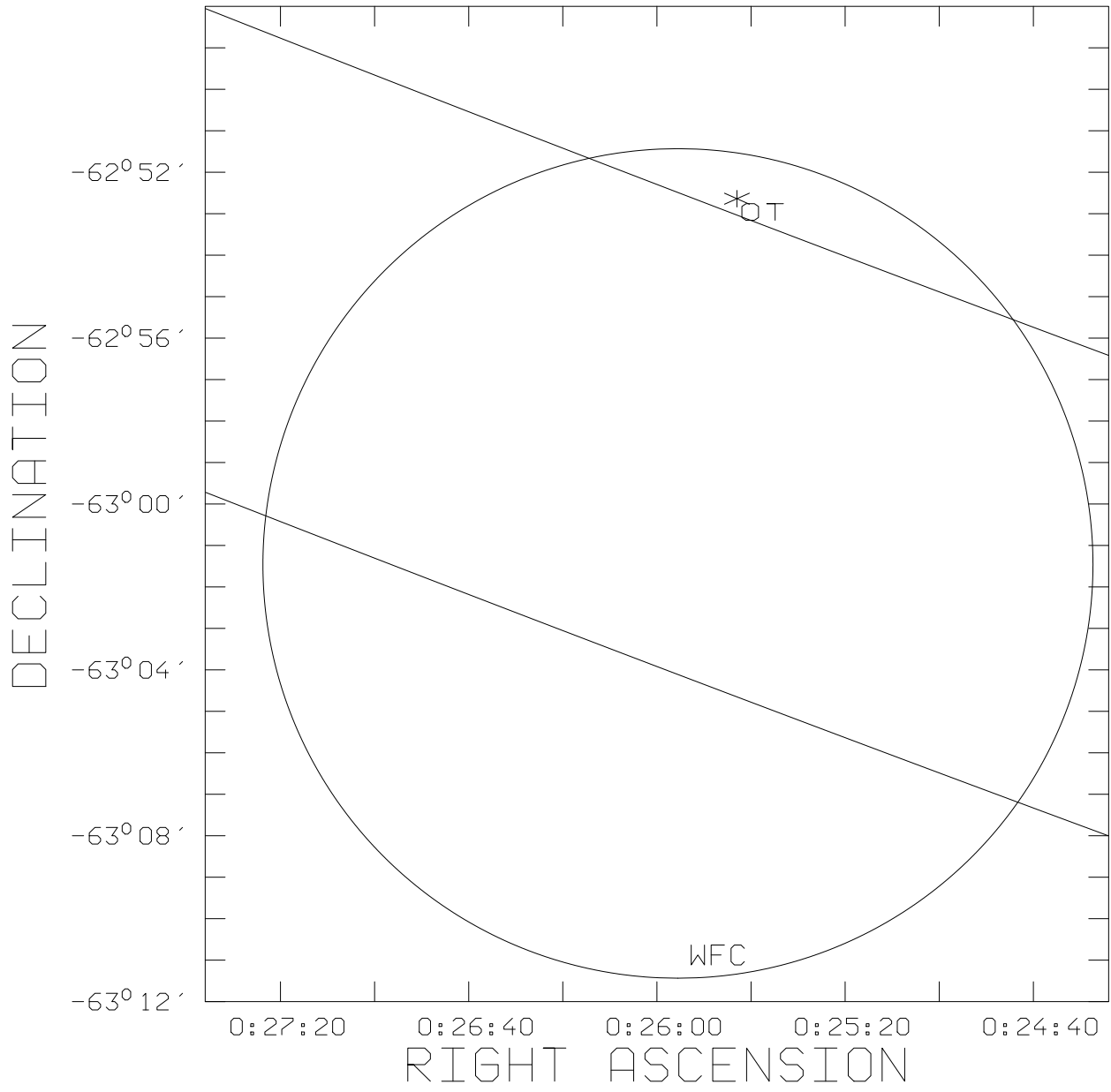
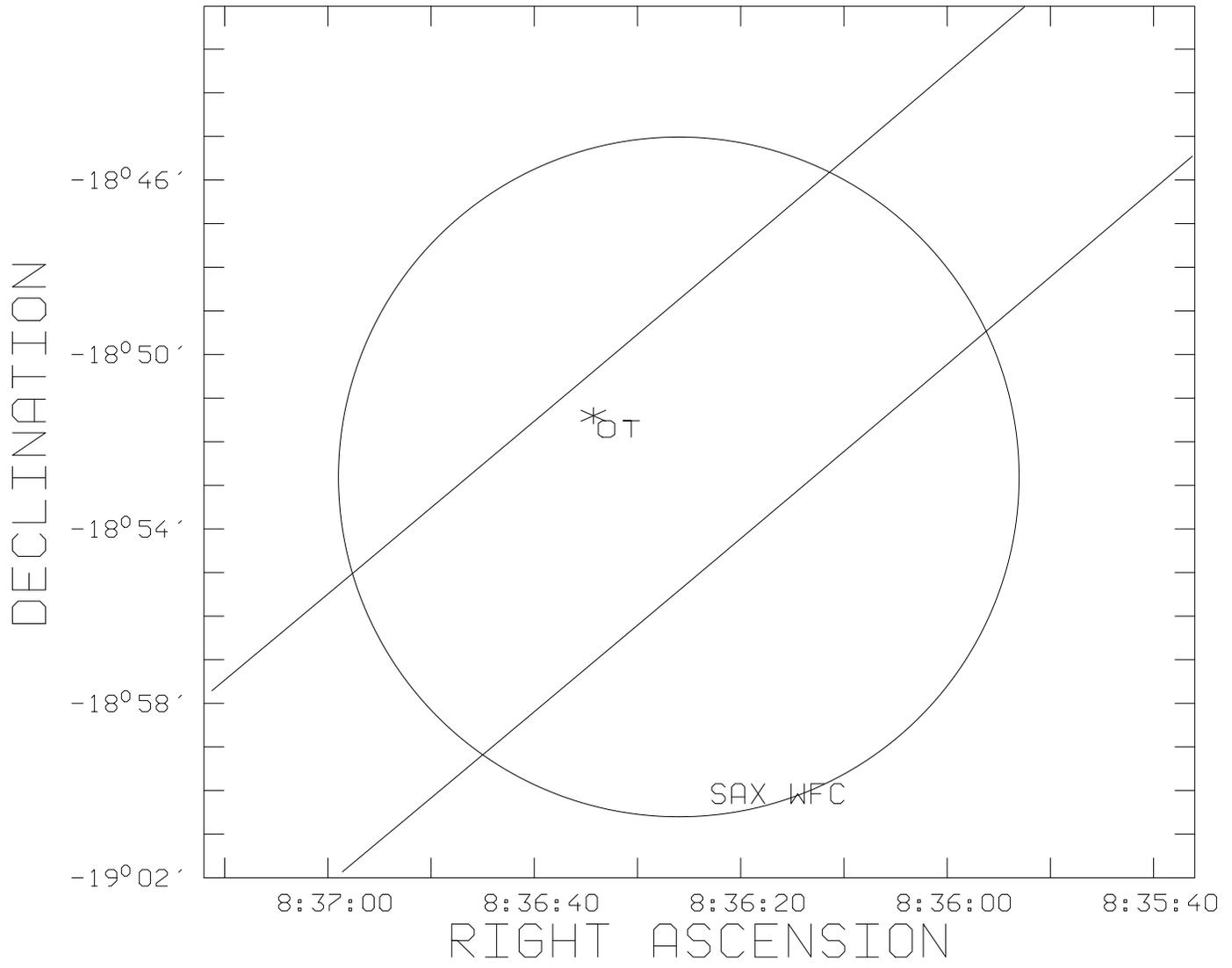
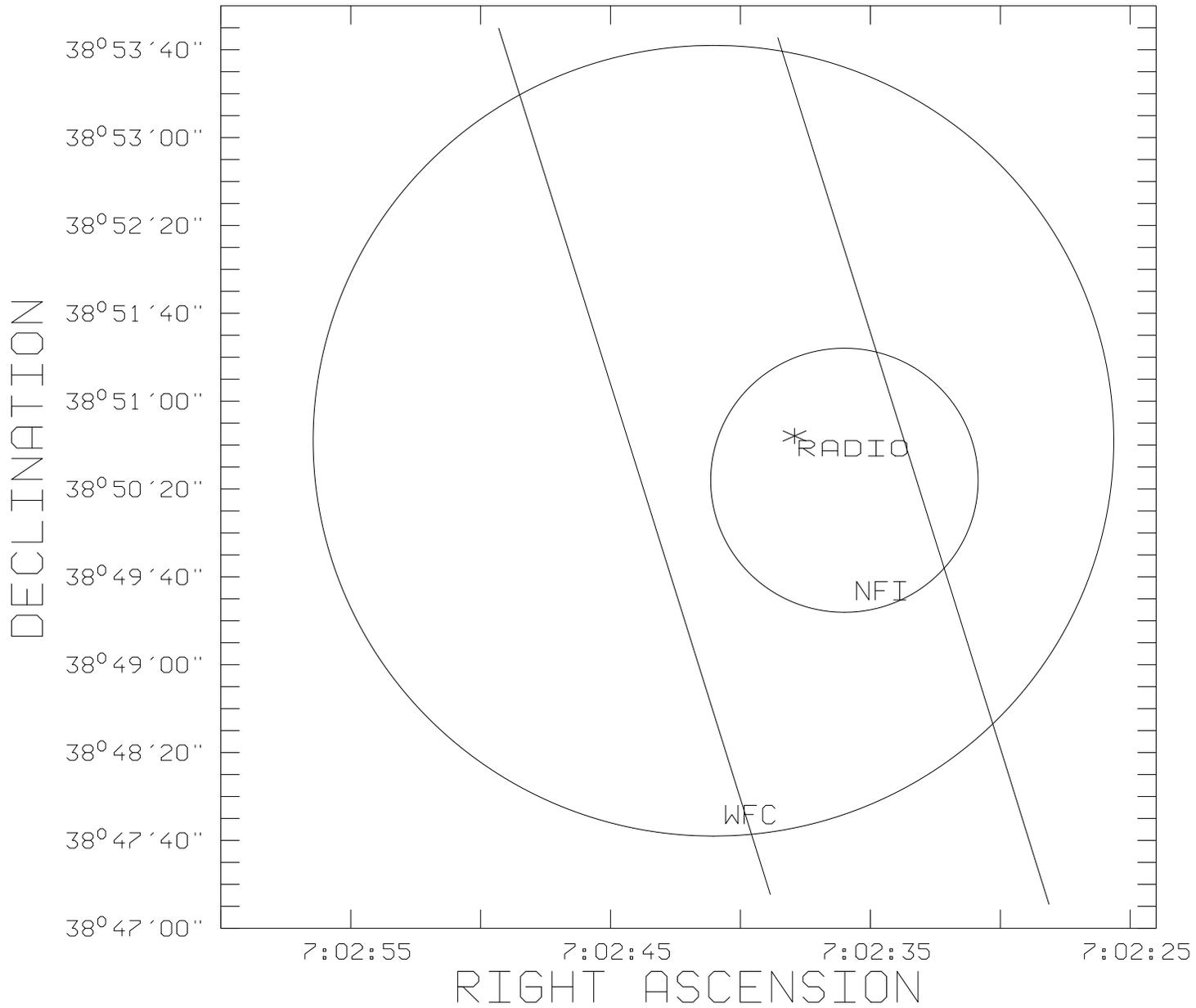
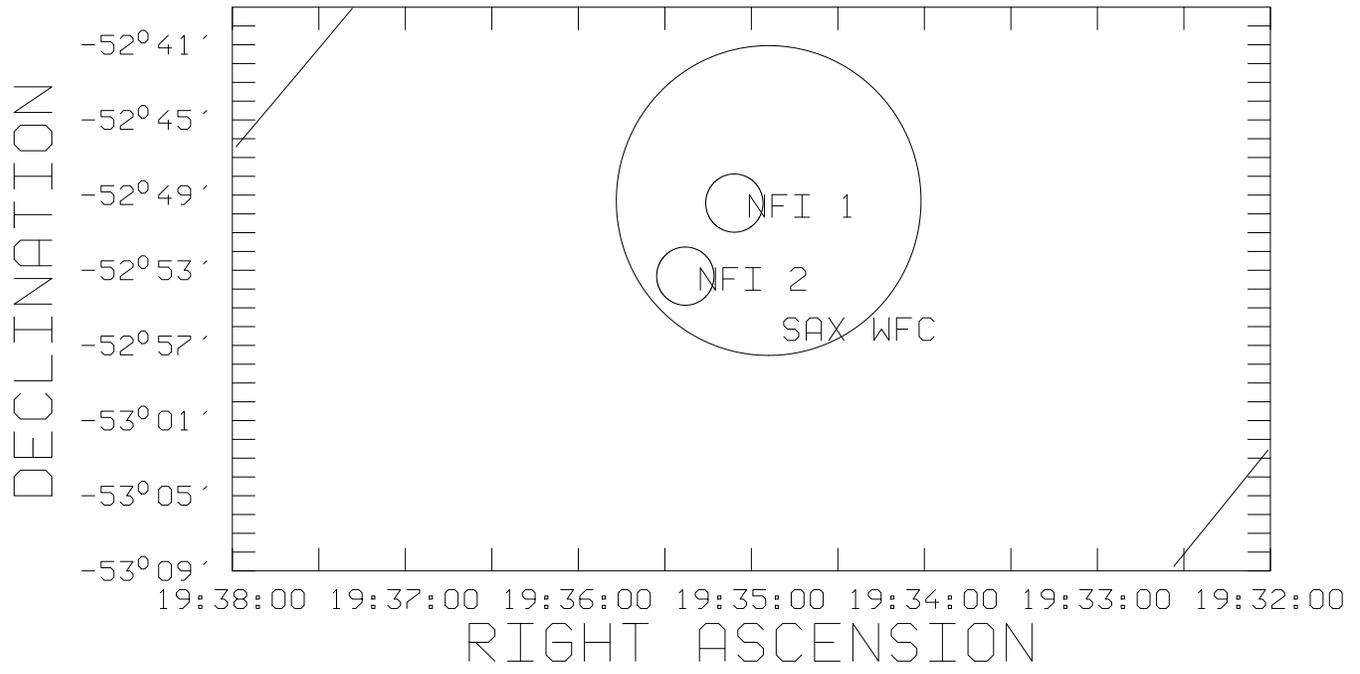
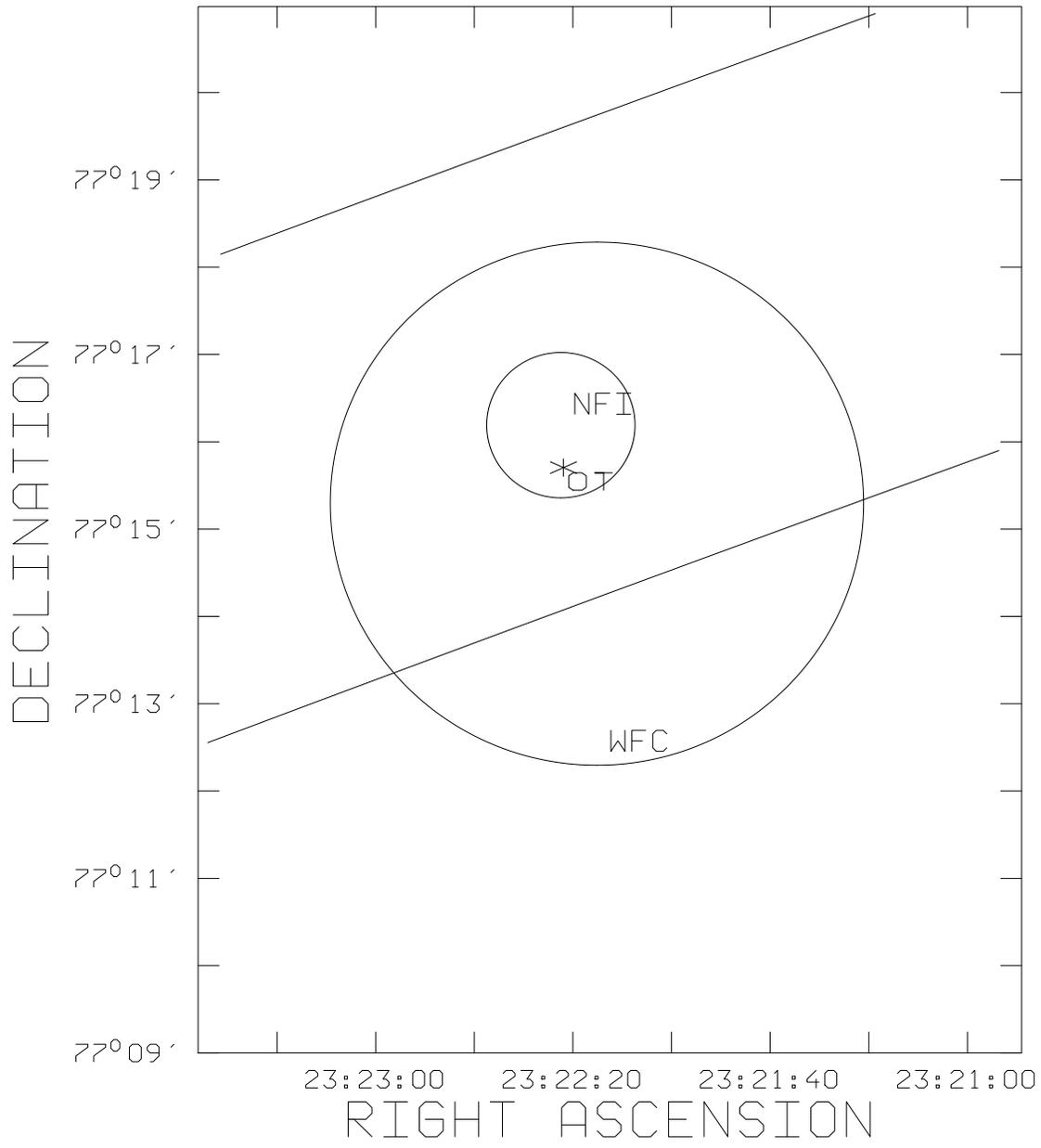
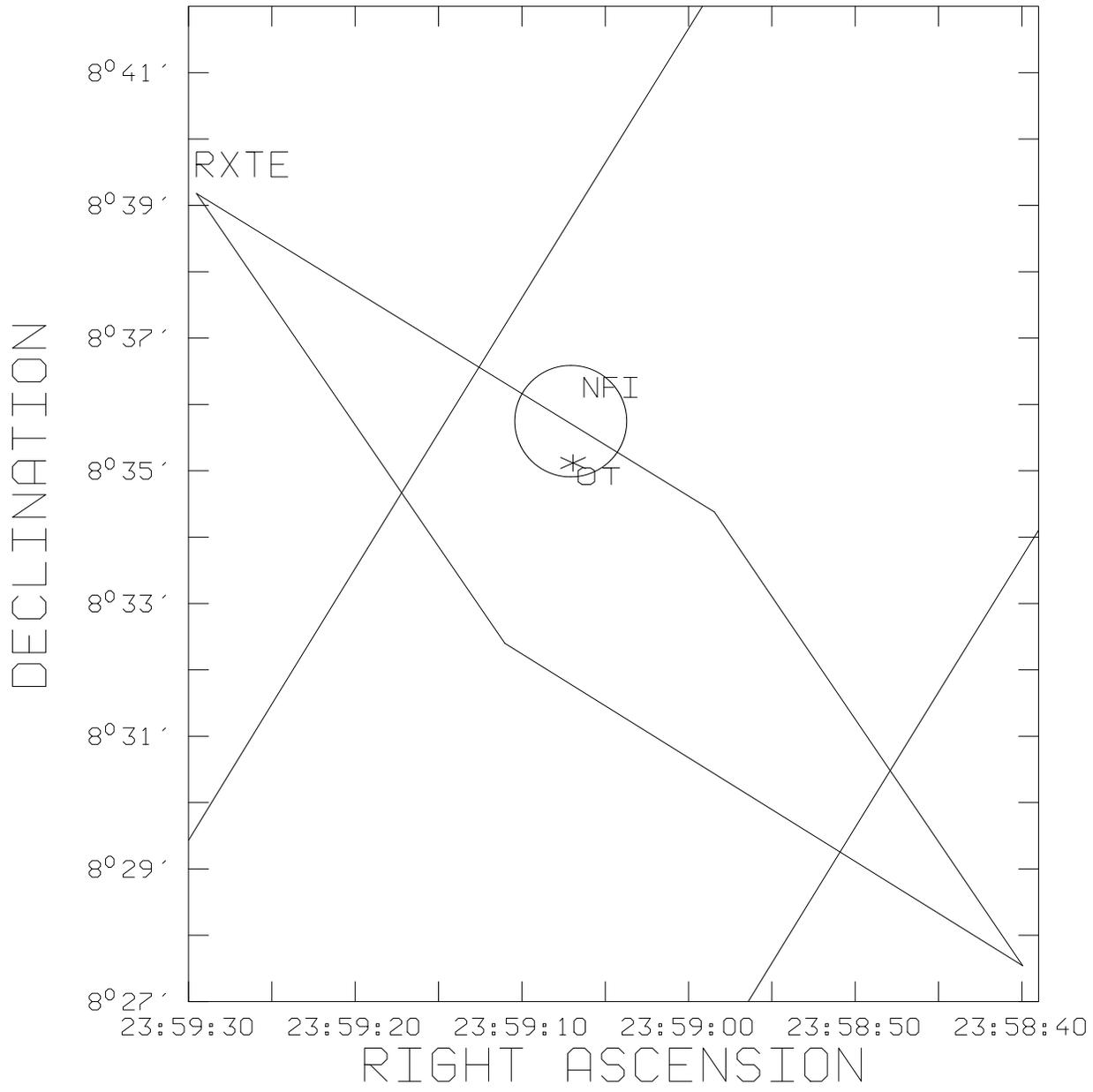
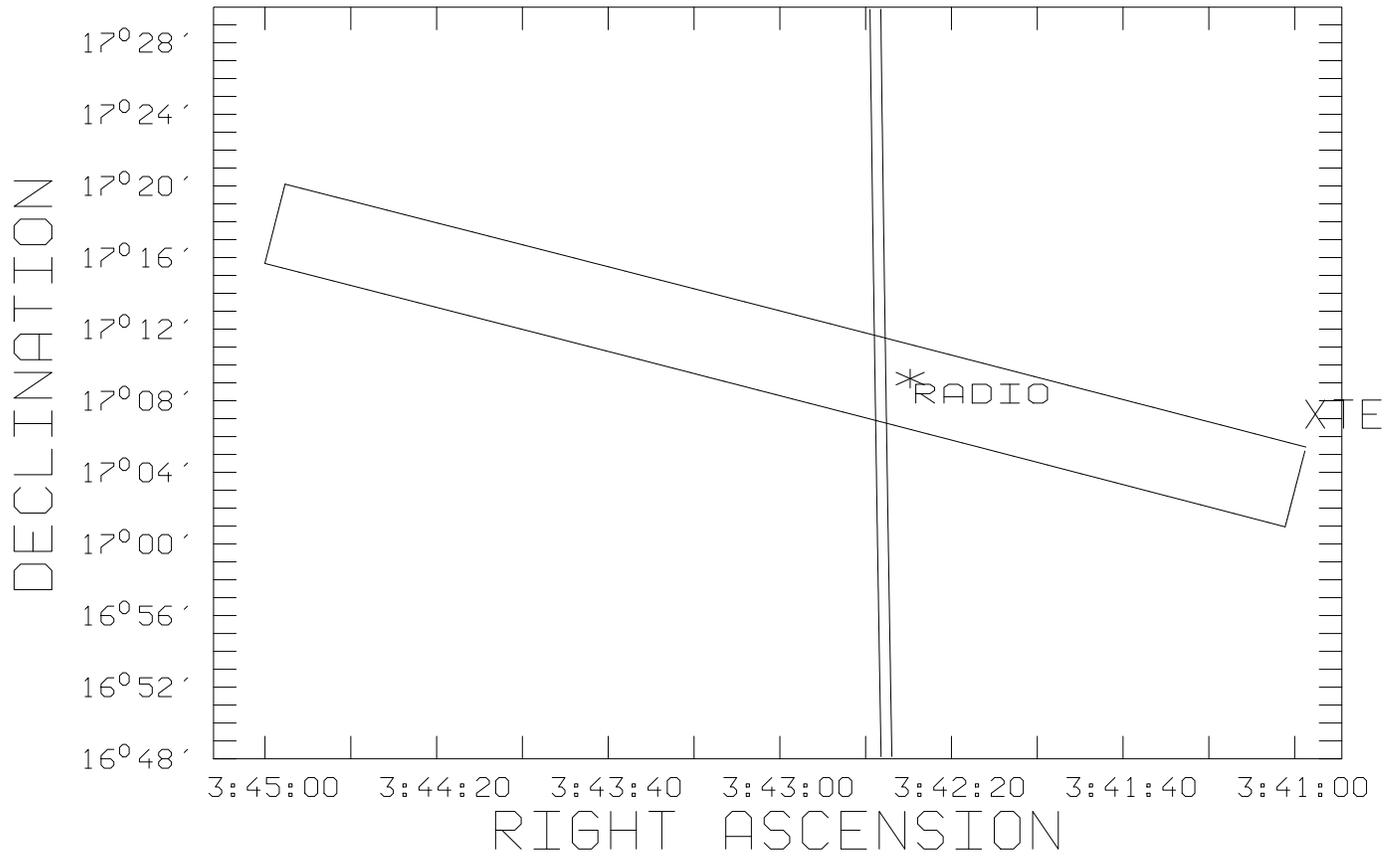
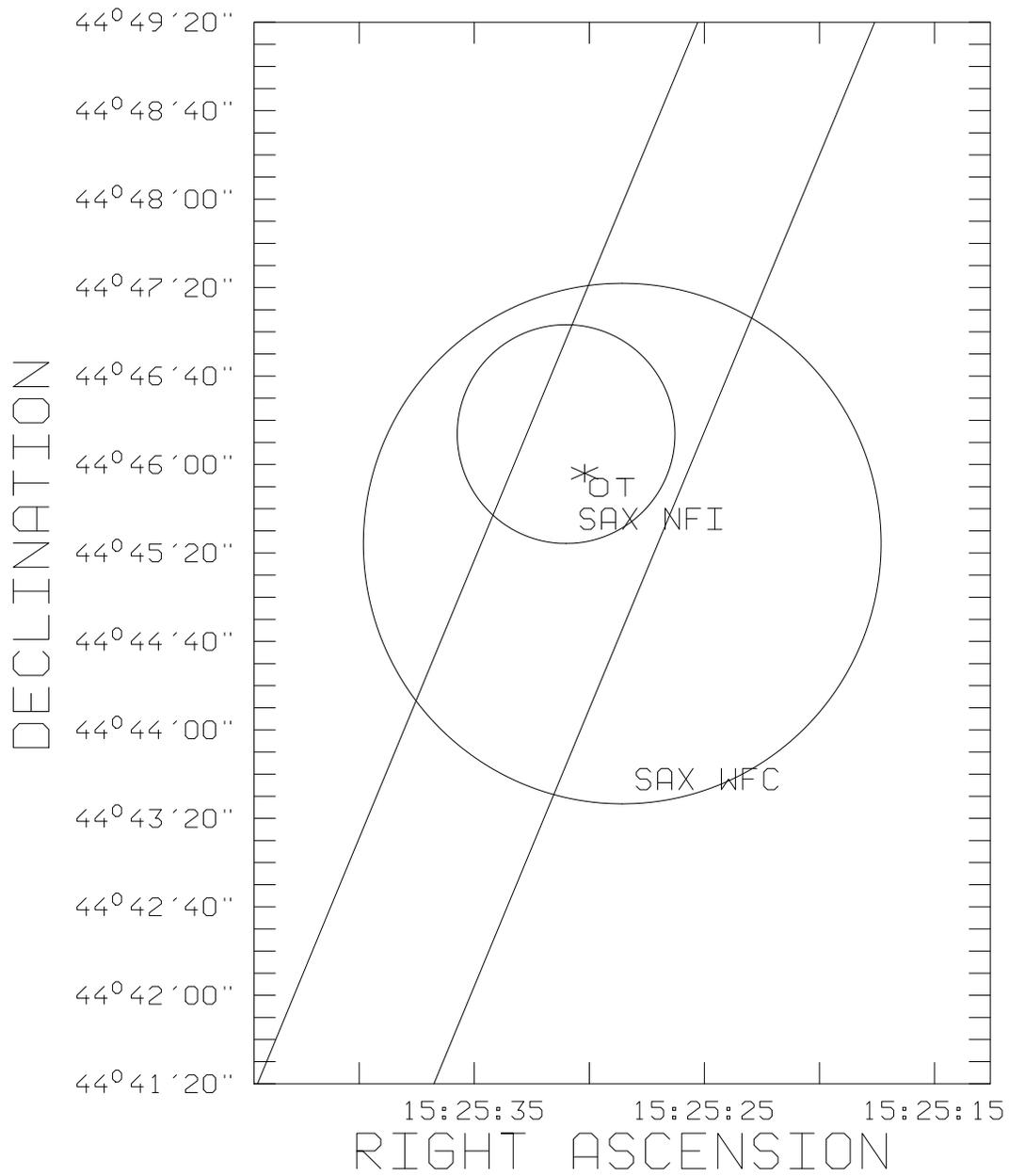
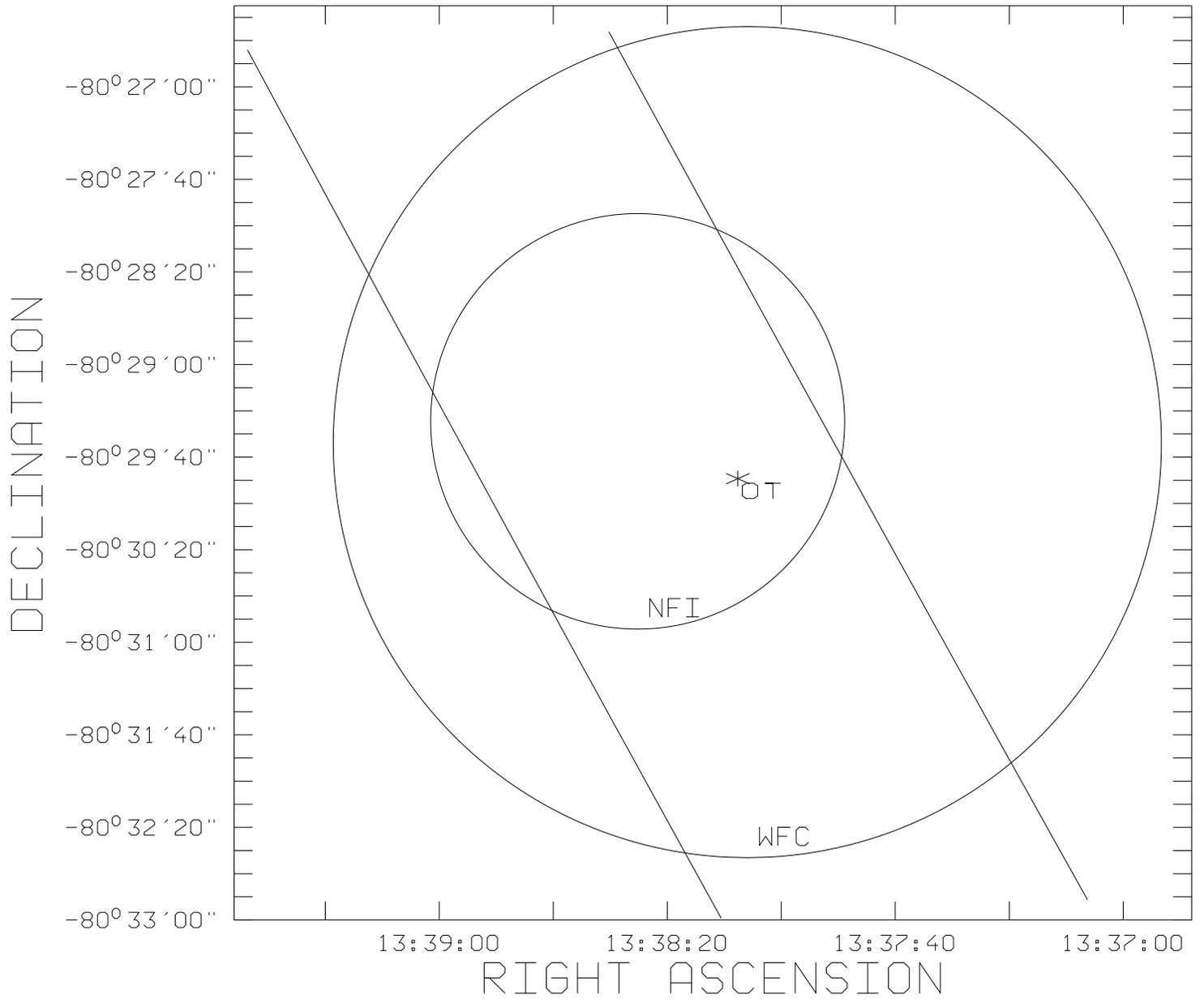
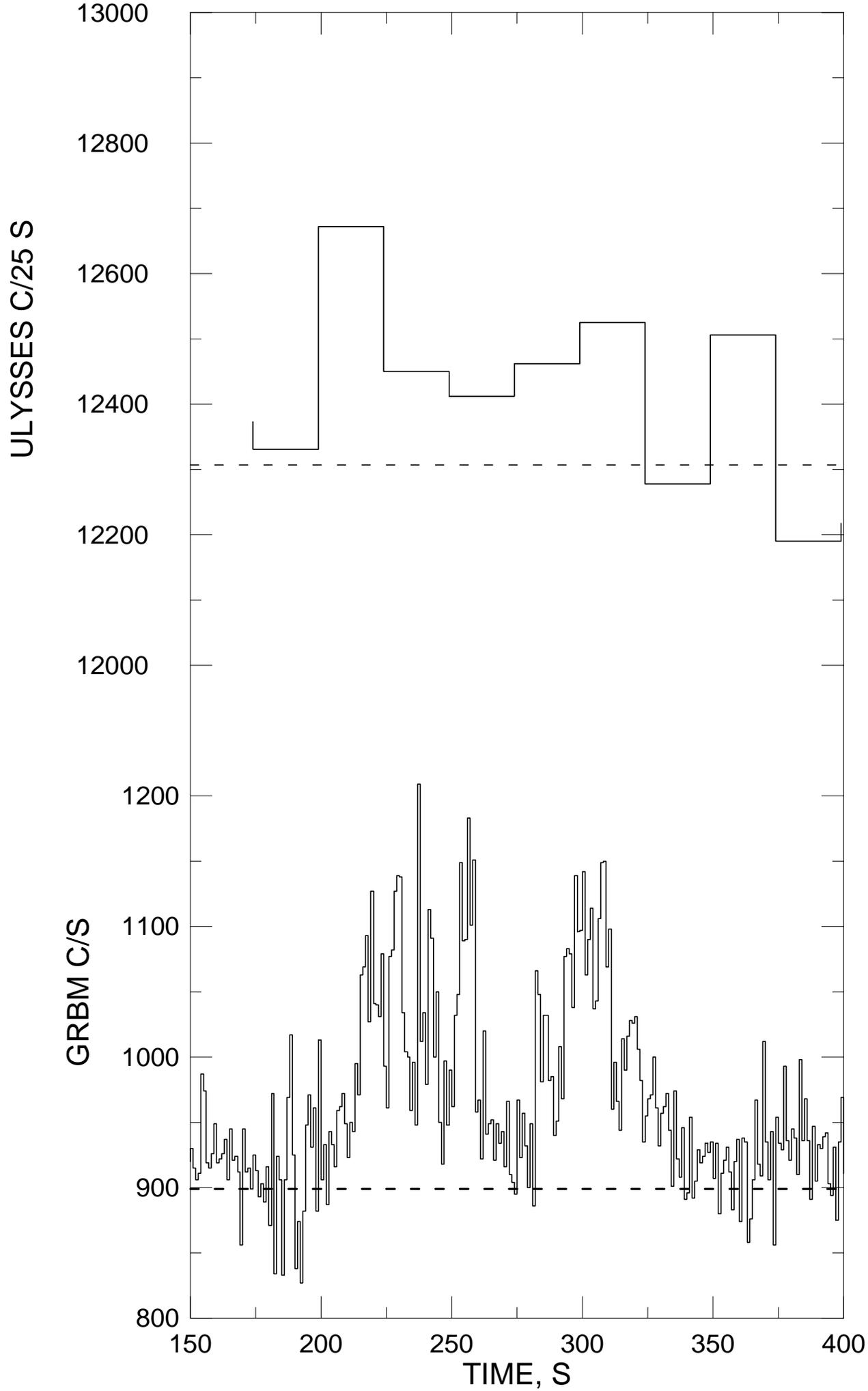
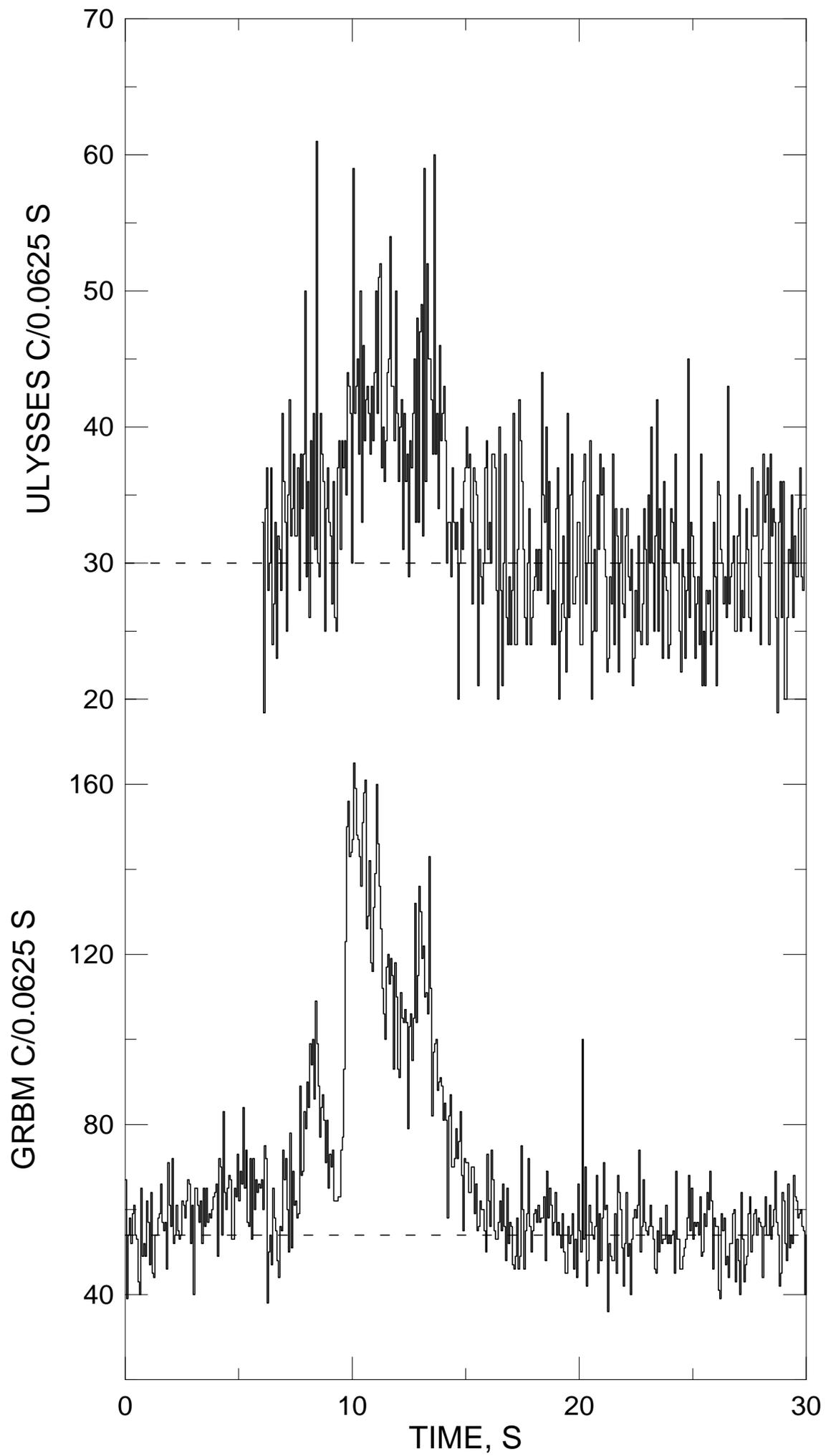
\end{document}